\documentclass[conference]{IEEEtran}
\IEEEoverridecommandlockouts

\usepackage{cite}
\usepackage{amsmath,amssymb,amsfonts}
\usepackage{algorithmic}
\usepackage{graphicx}
\usepackage{textcomp}
\usepackage{xcolor}
\usepackage{url}
\def\BibTeX{{\rm B\kern-.05em{\sc i\kern-.025em b}\kern-.08em
    T\kern-.1667em\lower.7ex\hbox{E}\kern-.125emX}}
\begin{document}

\title{The Immersive Archive: Archival Strategies for the Sensorama \& Sutherland HMD}
\author{\IEEEauthorblockN{Zeynep Abes}
\IEEEauthorblockA{\textit{Media Arts + Practice} \\
\textit{University of Southern California}\\
Los Angeles, USA \\
zabes@usc.edu \\
0009-0006-3928-0099}
\and
\IEEEauthorblockN{Nathan Fairchild
}
\IEEEauthorblockA{\textit{Mobile \& Environmental Media Lab
} \\
\textit{University of Southern California
}\\
Los Angeles, USA \\
nfairchi@usc.edu \\
0009-0004-8795-5180}
\and
\IEEEauthorblockN{Spencer Lin }
\IEEEauthorblockA{\textit{Mobile \& Environmental Media Lab} \\
\textit{University of Southern California}\\
Los Angeles, USA \\
linspenc@usc.edu \\
0009-0007-0974-3395}
\and
\IEEEauthorblockN{Michael Wahba}
\IEEEauthorblockA{\textit{Mobile \& Environmental Media Lab} \\
\textit{University of Southern California}\\
Calgary, Canada \\
wahbam@usc.edu \\
0000-0003-3586-9171}
\and
\IEEEauthorblockN{Katrina Xiao}
\IEEEauthorblockA{\textit{Mobile \& Environmental Media Lab} \\
\textit{University of Southern California}\\
Los Angeles, USA  \\
katrinax@usc.edu \\
0009-0003-2168-937}
\and
\IEEEauthorblockN{Scott S. Fisher}
\IEEEauthorblockA{\textit{Mobile \& Environmental Media Lab} \\
\textit{University of Southern California}\\
Los Angeles, USA \\
scottfis@usc.edu \\
0009-0005-8772-7433}
}
\maketitle

\begin{abstract}
The Immersive Archive is an initiative dedicated to preserve and restore the groundbreaking works from across Extended Reality (XR) history. Originating at the University of Southern California's Mobile and Environmental Media Lab, this archive is committed to developing and exhibiting simulations of influential XR devices that have shaped immersive media over time. This paper examines the challenges and strategies involved in archiving seminal XR technologies, with a focus on Morton Heilig’s Sensorama and Ivan Sutherland’s Head-Mounted Display. As pioneering prototypes in virtual and augmented reality, these devices provide valuable insights into the evolution of immersive media, highlighting both technological innovation and sensory experimentation. Through collaborative archival efforts with institutions such as the HMH Moving Image Archive at University of Southern California and the Computer History Museum, this research integrates media archaeology with digital preservation techniques. Emphasis is placed on documentation practices, restoration of physical artifacts and developing simulations of these historic experiences for contemporary virtual reality platforms. Our interdisciplinary approach to archival methodologies, which captures the multisensory and interactive qualities of these pioneering devices, has been instrumental in developing a framework for future immersive media preservation initiatives. By preserving the immersive essence of these early experiences, we lay the groundwork for future generations to explore and learn from the origins of immersive media. Safeguarding this rich legacy is essential to ensure these visionary works continue to inspire and shape the future of media landscapes.
\end{abstract}


\section{Introduction}

While extended reality (XR) technologies are still in the early stages of consumer adoption, these technologies have been around for decades, with a history usually traced back to the early 1960s.\cite{rheingold1991virtual} Many of the innovations made throughout the latter half of the 20th century established the core concepts and approaches that still define the field today. \cite{messeri2024unreal}Histories of XR typically either begin with a discussion of Morton Heilig’s Sensorama device or Ivan Sutherland’s head-mounted display (HMD).  Heilig’s work took the form of the construction of a multi-sensory, fully immersive cinema device called the Sensorama (1960) that influenced many of the pioneers of virtual reality (VR), most of whom then used computers to start building the future that Heilig envisioned.  Later on in this decade (1968), Sutherland pioneered an augmented reality (AR) tracking system and display, which marked XR’s first six-degree-of-freedom (6DoF) system.  Heilig’s vision of immersion and Sutherland’s techniques mark the beginning of a continuum of influence and technological development that bring us to the XR of today.  

We know that the history of XR is vastly under-documented. Even though it has been a dynamic medium that has been a part of our mainstream culture for decades, the current narrative surrounding XR’s history tends to focus on surface level interpretations that neglect its deeper technical, theoretical, and artistic background. These widely accepted timelines tend to only celebrate successful commercial applications instead of exploring the medium's conceptual foundations and creative experiments. Much of this medium's history is interconnected with the trajectories of technologies like computer science, interactive media, video games and cinema. \cite{soccini2017kusama} By exploring these intersections, we aim to illuminate how past innovations have shaped today’s media landscape and continue to inform its evolution. One of our main goals as an archive is to treat these early devices as valuable objects of study and analyze their impact on society, culture, and future technologies.  

In popular culture, Morton Heilig’s Sensorama is often celebrated as the first Virtual Reality device. Yet, Heilig’s contributions extend far beyond this single innovation. His extensive journals reveal a deep engagement with immersive storytelling, shaped profoundly by his experiences with the Cinerama during the 1950’s \cite{heilig_idea_books_2024}. Driven to push the boundaries of audience engagement, Heilig explored multi-sensory, 3D film techniques that immersed viewers in unprecedented ways. His work provides a media-theoretical lens through which we can trace cinema’s impact on the evolution of XR. Today, filmmakers creating live-action 360- and 180-degree experiences draw on these immersive storytelling techniques—often without realizing the legacy they are building upon. At the Immersive Archive, we have uncovered a fascinating historical record of Heilig's  projects through meticulous documentation of his journals and various physical artifacts, including correspondence, screenplays, storyboards, catalogs, purchase records, and more.

While Sensorama pioneered immersive storytelling methods for  audiences, Ivan Sutherland’s HMD advanced this concept by introducing interactivity, allowing users to explore virtual spaces that responded to their movements. This shift from observing to interacting marked a new chapter in immersive technology, setting the stage for modern XR. \cite{rheingold1991virtual}Ivan Sutherland who is often called the "father of computer graphics" created his head mounted display at his Harvard research lab that was a part of his many experiments and research on interactive computing. He called it the "head-mounted display" which was possibly the first recorded use of the term "HMD." The HMD is typically and inaccurately known as “Sword of Damocles” due to the mechanical system that supported the headset. Calling the device “Sword of Damocles” is another common misconception that keeps getting repeated within narratives surrounding XR history. The HMD provided a visual overlay of digital shapes that responded to head movements which integrated various fields within computer science, notably real-time processing, computer graphics, and human-computer interaction. At a time when computers had limited use, Sutherland's work emphasized interactive computing, which became fundamental in the development of computer graphics and immersive systems. 

The historical significance of these early works cannot be overstated; however, our perception remains limited due to poor preservation and restricted access to their detailed history. the Immersive Archive combines a multifaceted approach including the creation of a participatory digital archive, in-depth media archaeological research as well as simulations of these seminal creations, all intended to illuminate the significance of such immersive works within the contemporary media landscapes. XR is not “new” media. Our current concept of virtual, augmented and mixed reality has been around for over half a century and stems from a murky history of conflicting narratives and inflated expectations. People often focus on the “new" and the dominant technologies, which are driven by institutional narratives - we would like to uncover what has been left out and examine the margins. Reflecting on a critical perspective of "new" media, we aim to deconstruct the history of XR and take an alternative approach to the conventional narrative of its development. 

\section{Resurrecting The Sensorama: An Immersive Multi-Sensory Experience}

Our work to bring Sensorama back to life began with USC’s acquisition of the Sensorama device itself.  After Heilig’s death in 1997, his family attempted to sell his archival materials – including his physical Sensorama devices and original film strips – to museums around the world.  At this time, XR was entering a market trough and a cross-disciplinary wane in interest.  The desired deal between a major museum and Heilig’s family never materialized. In 2017, the Heilig family reached out to Professor Scott Fisher at USC, a VR, practitioner and former acquaintance of Heilig, for assistance in finding a home for these materials.  Fisher secured a place for Heilig’s complete archival materials in the HMH Moving Image Archive at USC.  These materials included all of Heilig’s hardware, artifacts, his original 3D films, and his personal journals. 

At the time of this acquisition, Heilig’s artifacts were in a severe state of disrepair.  For years, his original hardware had been housed in a storage unit with roofing issues, leading to water damage. The Sensorama devices were highly complex pieces of machinery that delivered multisensory media experiences. Each machine included a film projector and optics for presenting stereoscopic films, an analog stereo audio system, a “smell bank” of scents with a smell delivery system, a fan system for wind simulation, and a haptic chair.  These devices were no longer functional, and the remaining film prints of Heilig’s immersive films had suffered from aging and exposure to the elements. Once the collection was transported to the HMH Archive and upon further inspection and research, additional challenges became clear, such as obsolescence of mechanical parts.  Since none of Heilig’s materials were digital, no options existed to purchase digital components for direct emulation or simulation. Together, these issues presented significant challenges to any archival restoration efforts.   

While much of the hardware restoration of Heilig’s devices is still ongoing at the HMH Archive under the leadership of archivist Dino Everett, our work as the Immersive Archive has focused on using Heilig’s materials to build a virtual simulation of using the Sensorama device. This work represents the Immersive Archive’s core approach to restoring immersive media work, which is: capturing and presenting the first-person experience of this work to audiences. 

\begin{figure}
    \centering
    \includegraphics[width=0.5\linewidth]{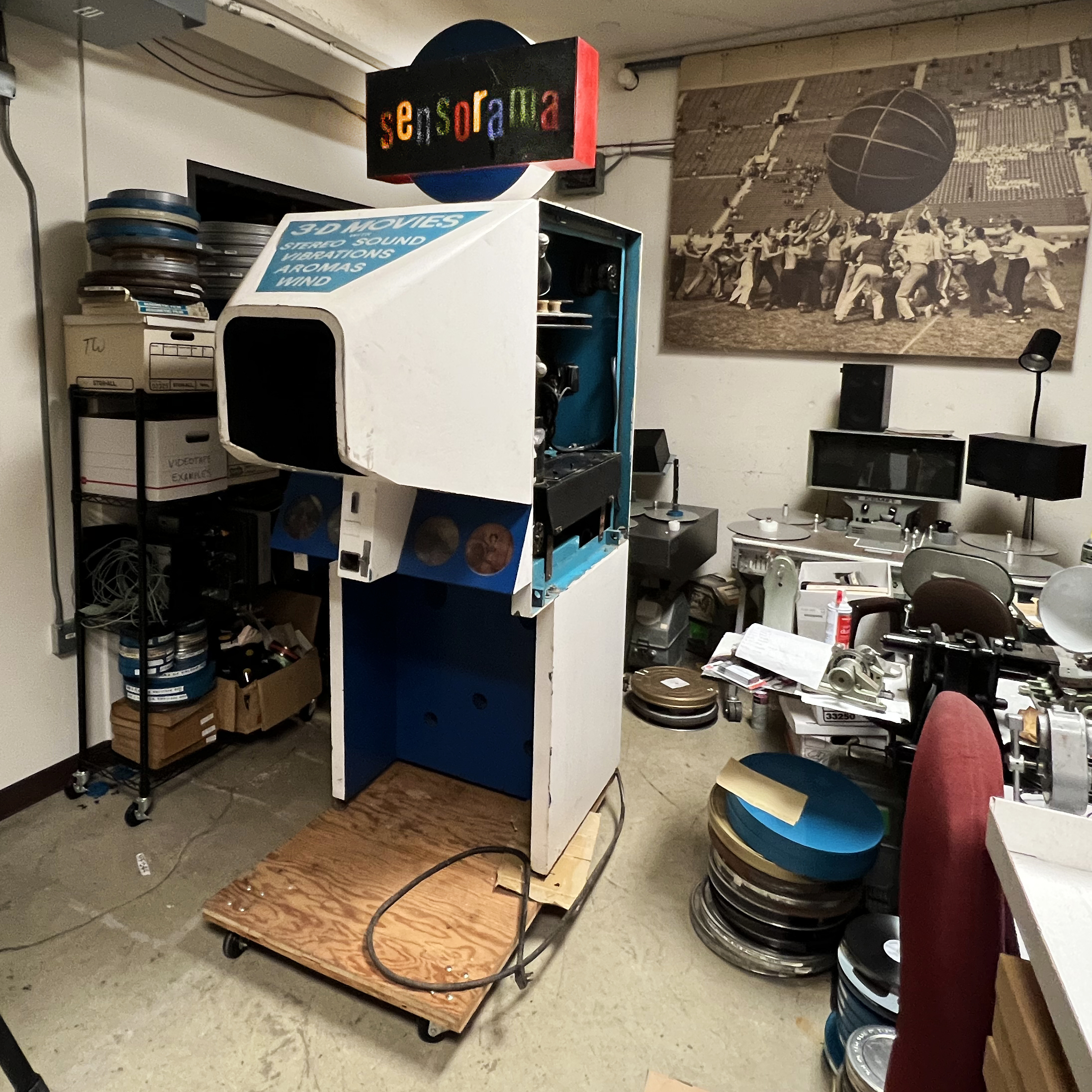}
    \caption{The Sensorama at the HMH Moving Image Archive}
    \label{fig:sensorama-hmh}
\end{figure}

This process started with using photogrammetry tools to do a detailed scan of one of Heilig’s Sensorama devices, so that audiences could view a detailed ‘digital twin’ of the Sensorama cabinet at scale in a VR headset.  For the scan, we removed the device’s side panels so that audiences could view the internal workings of the device.  In our Unity-based VR simulation of the Sensorama experience, this digital twin of the Sensorama sits in a virtual gallery where users can examine the device while archival video of Heilig showing off the device plays on the virtual gallery wall.  

The next essential item for us was digitizing Heilig’s stereoscopic film prints so that viewers of our archive could experience the original Sensorama films.  Each of Heilig’s films for Sensorama was shot using two side-by-side 16mm cameras, with the imagery from each camera printed as an image pair onto a single strip of 35mm film for projection into the Sensorama’s viewer optics. The stereo audio was recorded separately and then synchronized on the final strip. The HMH Archive has worked to restore Heilig’s film prints and digitize each eye at 5K resolution.  Our team took these videos and used Adobe After Effects to create a stereoscopic video file for viewing in contemporary VR headsets.  In our virtual Sensorama experience, when viewers walk up to our 3D scan of the Sensorama device and line their eyes up to the device’s optics, the scene transitions to an immersive video player that presents each Sensorama video at their initial 120° field of view. 

In this manner, our team has developed an experience in which any viewer with a VR headset can see and hear the remaining archival footage of Heilig himself, examine a digital twin of the Sensorama device in detail, and have the audio-visual experience of viewing the original Sensorama films.  The clear missing element of our virtual restoration of the Sensorama is a lack of haptics and smell.  This highlights both how ambitiously multisensory Heilig’s work was, and the difficulty of archiving and distributing immersive experiences that contain more sensory elements than sheerly sight and sound. 
\begin{figure}
    \centering
    \includegraphics[width=1\linewidth]{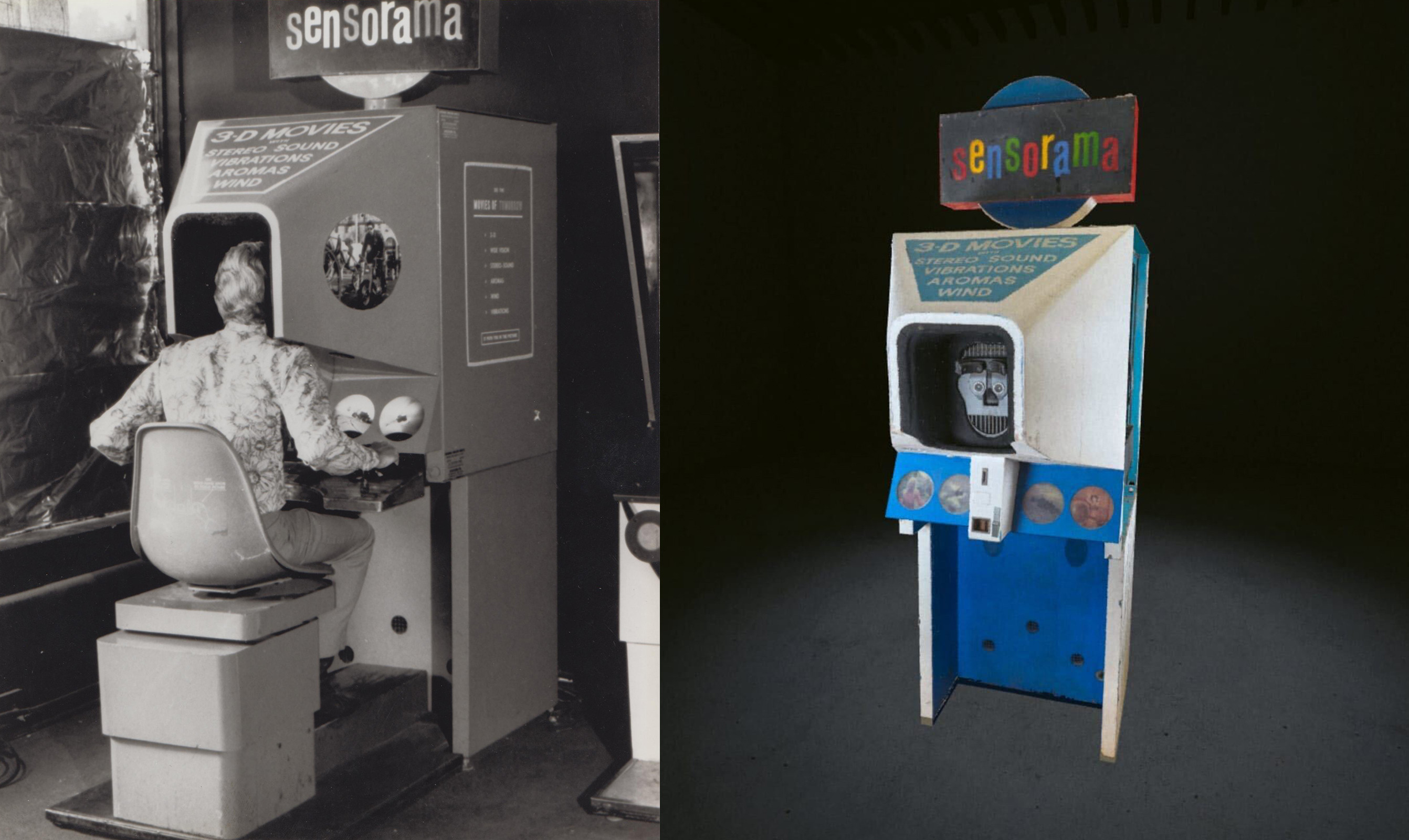}
    \caption{Original archival photo of the Sensorama (left) alongside our 3D simulation viewable in a modern VR headset (right)}
    \label{fig:sensorama-comparison}
\end{figure}

\subsection{Chronicling Heilig's Journals}

Apart from the Sensorama, we also have access to various items from Heilig's collection of artifacts. The most notable of these are his “Idea Books” that he had kept up from 1943 until the 1980’s. His journals are highly personal and reflect his eccentric personality and inventive approach to technologies. Our team extensively delved into Heilig's journals to chronicle his life and how his ideas around immersion eventually culminated to create the Sensorama. 

His first journal entries in 1943, at age 17, reflects on his life in Long Beach, his dissatisfaction with Cornell's academic atmosphere, and his budding interest in cinema, alongside awareness of WWII. His constant commentary on how to improve society using art and technology is apparent throughout his idea books, as his passion for cinema flourishes. 

Heilig was fascinated with the Cinerama and its immersive nature, noting how its sensory impact draws viewers into a heightened state of awareness, allowing for deeper observation and memory retention. This is a stepping stone to his lifelong interest to envelop viewers in a cinematic experience.
\begin{figure}
    \centering
    \includegraphics[width=1\linewidth]{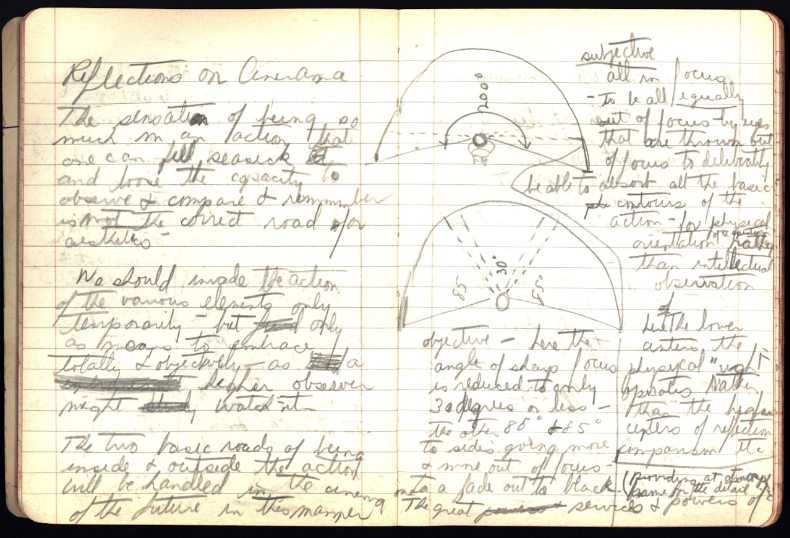}
    \caption{Idea Book 21, 1954}
    \label{fig:idea-book-21}
\end{figure}
“Human perception exists with a certain range of dimension. It is theoretically possible to create complete sensory robots on a variety of dimensions beyond those of normal man – so that he can morally project himself and physically react with environments that are completely foreign to him.” (Idea Book 26, 1957)
\begin{figure}
    \centering
    \includegraphics[width=1\linewidth]{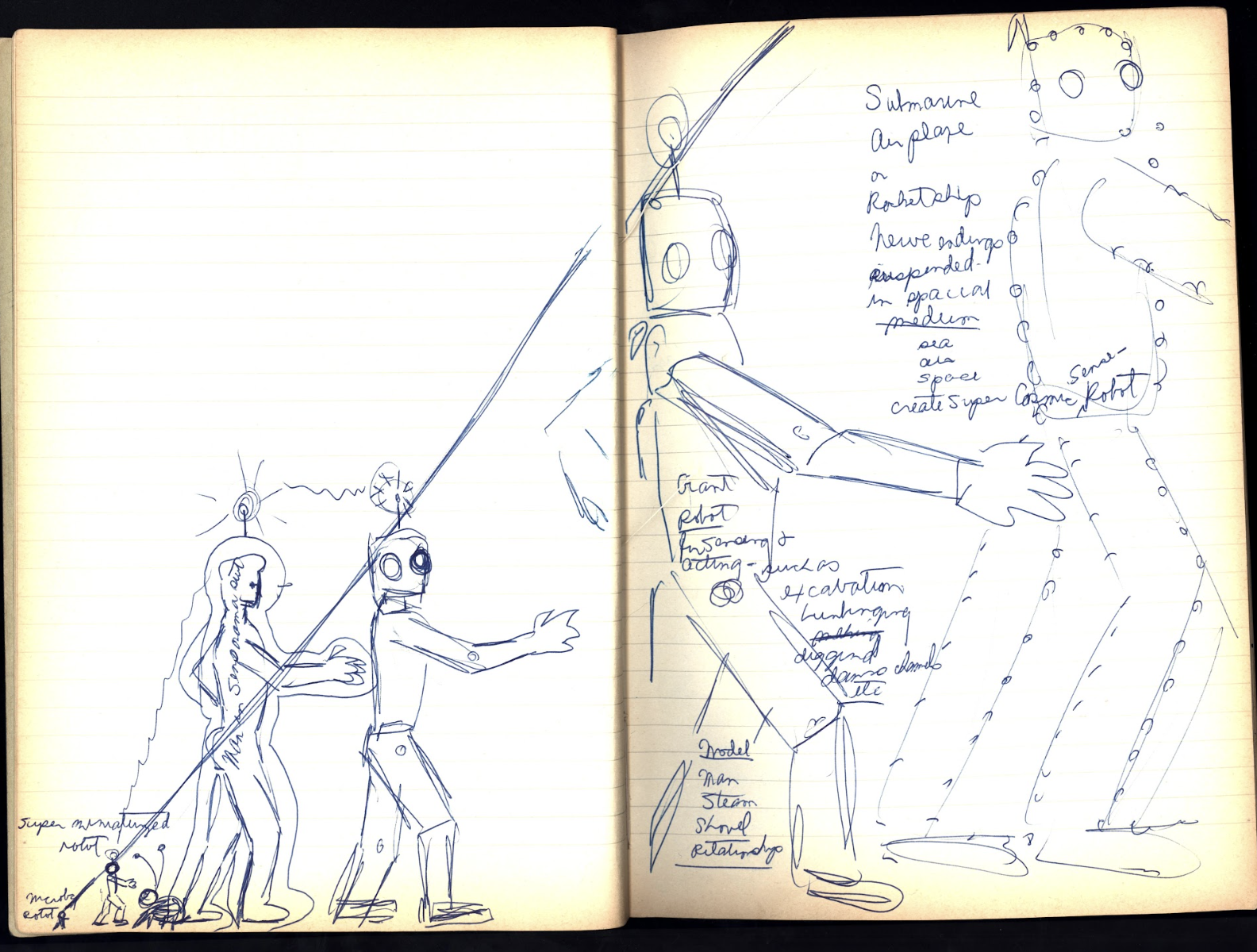}
    \caption{(Idea Book 17, 1952)}
    \label{fig:idea-book-17}
\end{figure}
Heilig also has random everyday inventions throughout the books spanning from virtual reality concepts to practical everyday inventions. His restless curiosity across disciplines is evident as his wide-ranging imagination to improve human experiences through technology and sensory immersion can be seen in his sketches and notes. The "Sleep Machine" entry, for instance, reflects Heilig’s visionary approach to creating environments that influence states of consciousness—something he seemed deeply interested in, whether for relaxation, sensory immersion, or personal well-being. It's intriguing how he anticipated features like soundproofing, temperature control, and even gentle waking mechanisms that align with modern wellness technology.

Heilig’s ideas leap from one field to another, underscoring his ability to anticipate trends and needs that would only gain widespread attention decades later. His inventions, though seemingly random, embody a continuous drive to enhance human experience through technology and ergonomics, revealing a visionary whose ideas were far ahead of his time. \cite{10.1162/pres.1992.1.3.279}
\begin{figure}
    \centering
    \includegraphics[width=1\linewidth]{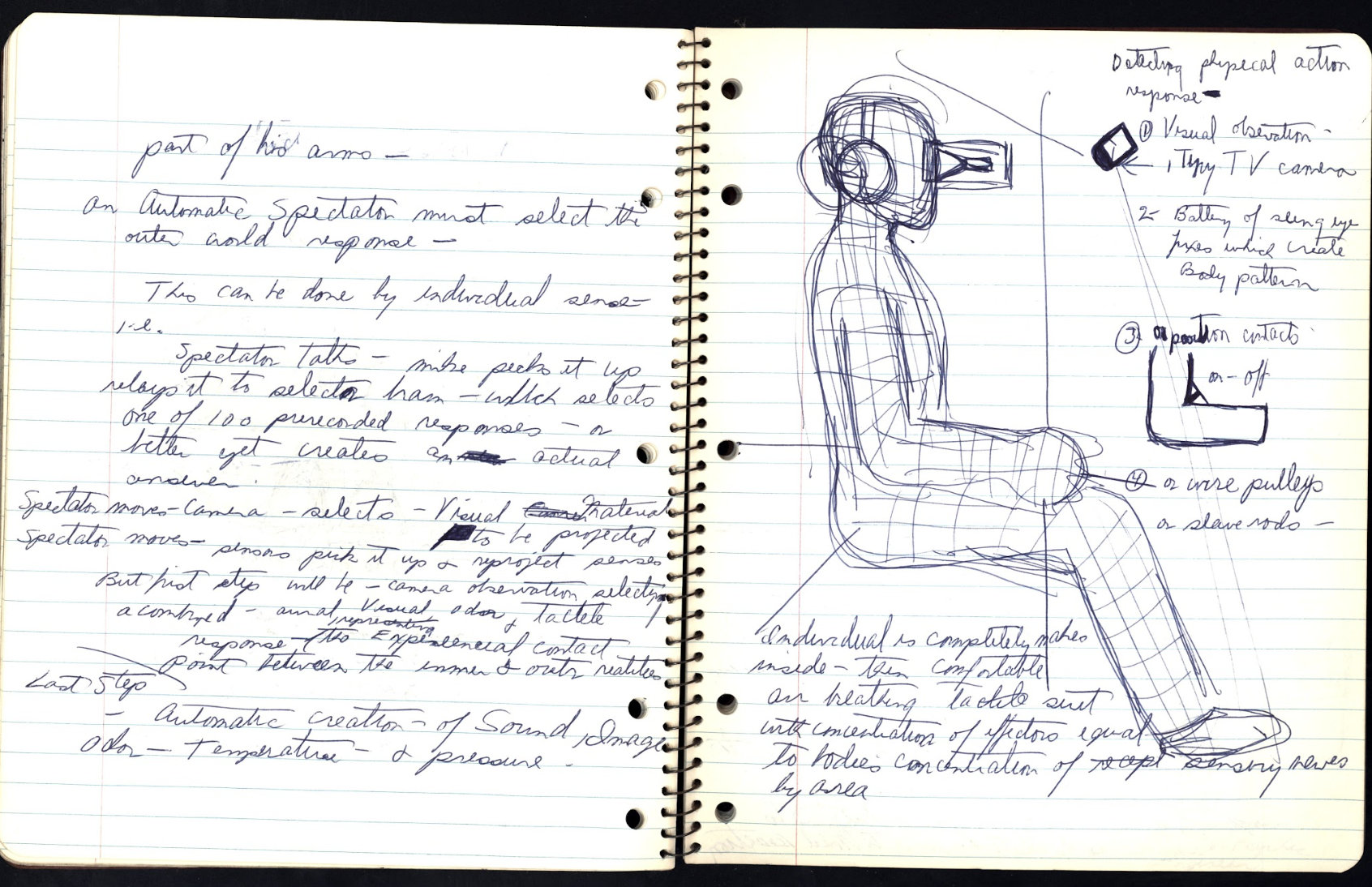}
    \caption{(Idea Book 26, 1957)}
    \label{fig:idea-book-26}
\end{figure}

\section{Sutherland HMD}

\subsection{Overview}
While Morton Heilig laid the foundation for immersive filmmaking, Ivan Sutherland and his team paved the way for XR as we know it today by developing the first head-mounted display (HMD) in 1968.

Our recreation of the Sutherland HMD experience puts users in Sutherland’s lab at Harvard and asks the user to try on the HMD. Users are free to walk around and inspect various period-accurate elements in the space, including a detailed photogrammetric scan of the actual HMD. An audio clip of Sutherland’s voice extracted from a talk he delivered at the Computer History Museum in March 19, 1996, “Virtual Reality Before it had a Name”\cite{VirtualRealityBeforeItHadThatName}, is played in the background to provide users with further context for the device. Finally, users are teleported into the HMD to experience what it was like to wear it while learning about the technical components of the device.

\subsection{Lab Environment Recreation}
The Harvard lab was recreated using reference images taken from several publications as well as media from the Computer History Museum (CHM) \cite{sutherlandhmd}\cite{VirtualRealityBeforeItHadThatName}\cite{ComputerHistoryMuseum}. In addition to the historical documents, a member of our team physically documented the scale dimensions and spatial layout of the room which used to contain Sutherland's lab at Harvard. 

We consulted former lab members, Bob Sproull and Harry Lewis  to capture nuanced details, such as the warning to students to refrain from using the HMD's mechanical sensor as a chin-up bar. These minute details enabled our team of 3D artists to meticulously model and texture the virtual lab and its furnishings, while staying as close to the real-world scale as possible. 

Representing the HMD itself as accurately as possible was a priority for our team. Members of our team traveled to the Computer History Museum in Mountain View, California to photogrammatrically scan the Sutherland HMD. After processing the scan and converting it to a 3D model, we included it in our recreation of Sutherland's lab.

\begin{figure}
    \centering
    \includegraphics[width=0.75\linewidth]{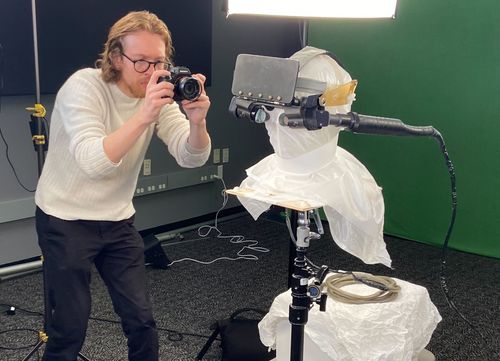}
    \caption{3D Scanning Sutherland's HMD}
    \label{fig:sutherland-hmd-scanning}
\end{figure}

\subsection{HMD View Recreation}
Recreating what it was like to look through the Sutherland HMD required the use of multiple post-processing effects within Unity. Knowing that our user would be using a modern VR headset, it was imperative to recreate the limited field-of-view and other aberrations present with the Sutherland HMD. A vignette effect was applied to lower the user’s field of view by occluding the user’s periphery with a dark shading. Then, a dust layer and chromatic aberration were added to simulate the optical imperfections in the HMD displays and lenses. We also added an ever-present CRT humming noise that a wearer would have experienced with the real headset. 

We present the viewer with recreations of the real demonstrations developed to show the capabilities of the HMD. One demonstration showed a vector render of a cyclo-hexane molecule showing potential XR applications in biochemistry. Another demonstration shows a room with in-laid letters meant to exhibit the parallax and precision that come with 6 degrees of freedom in a 3D digital environment.\cite{sutherlandhmd}. These demonstrations are not rendered in true vector graphics within the Unity game engine but are using shaders to exhibit how vector graphics in the headset would have appeared. 

\begin{figure}
    \centering
    \includegraphics[width=1\linewidth]{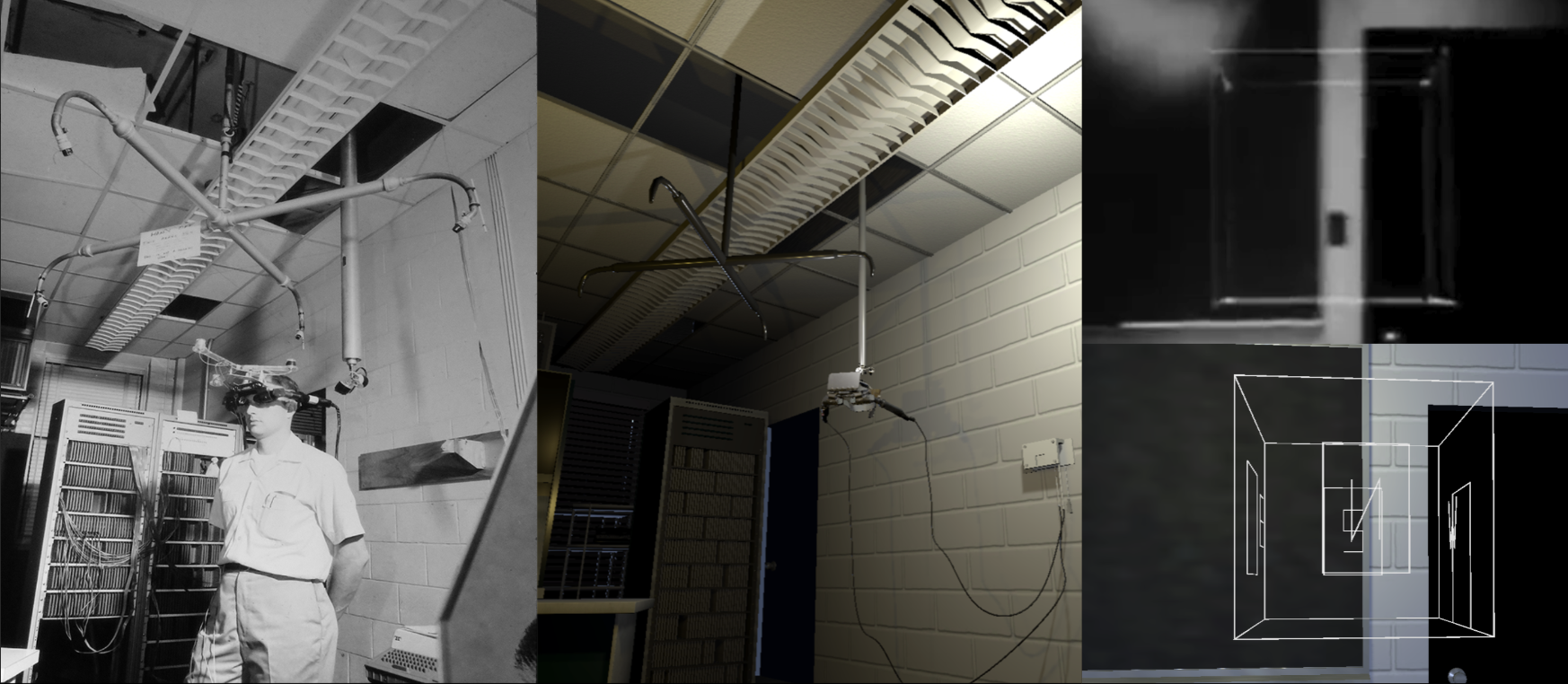}
    \caption{Original archival image of the Sutherland Head-Mounted Display setup (left), alongside our 3D simulation of the lab and hardware (middle);  on the top-right is the only existing camera-captured imagery of the original vector graphics seen through the headset, while the bottom-right shows our recreation of this viewpoint with a later version of the vector asset that was most commonly shown to viewers in the original experience.}
    \label{fig:sutherland-hmd-comparison}
\end{figure}

\section{Discussion: Limitations and Feedback}

While we feel that the core strength of our initiative is allowing present and future generations to get an experiential sense of XR’s early works, we are conscious of the project’s limitations. When it comes to restoring software-based experiences like Sutherland’s, we cannot consider our work restoration in the same manner that Heilig’s film clips are being restored.  Rather, with work like Sutherland’s – software that ran on bespoke hardware – we are limited to making a simulation of the experience using completely different, contemporary tools. While our work aims to capture the original experience as closely as possible using available imagery, interviews, and other documentation, our experience will always differ from the original work. Additionally, in the area of representing the tactile experience of any of these works – the feeling of the original hardware and other factors – our simulations cannot capture the full tangible experience that users would have had experiencing these works at the time. 

Our first public demonstration of the Immersive Archive was at SIGGRAPH 2023, where the team received feedback from a variety of users.  One consistent point of feedback from university professors was excitement that at last they could demonstrate to students these same early XR works that previously were confined to lecture slides. 
We also received a flood of feedback from many users about which works from the next stages of XR’s history – particularly the 1980s and 1990s – should be the following steps for our team to restore or simulate.  
Several experts in the field pointed out small details that could be modified for historical accuracy, which we used to modify our archival modules after the conference.   
Demonstrating the project again at the Augmented World Expo (AWE) 2024, which attracted a more strictly industry-focused demographic compared to SIGGRAPH, the audience was captivated by the opportunity to interact with seminal XR works. Many noted that they were unaware this kind of history existed, despite having worked in the XR industry for years. This kind of accessibility to XR’s history is precisely what we are trying to challenge, by not only preserving the history but also continuously studying it and keeping it alive through conferences, festivals, and exhibitions. To echo Wendy Chun: “Crucially, memory is an active process, not static. A memory must be held in order to keep it from moving or fading. Memory does not equal storage.”\cite{chun2008enduring} To preserve XR history, our team believes we must actively share archival work, ensuring visibility and engagement to keep this history alive.

\section{Connecting Archives}

Outside of our initiative, there is no centralized archive of XR history and media – no publicly accessible hub where XR content from across its history (which dates back to the 1960s) can be viewed and experienced. Instead, to the extent that this work is accessible at all, each of the medium’s key projects exists in a disparate location – scattered across various different museums, institutions, and personal collections. the Immersive Archive seeks to bridge isolated collections by creating a central archive of immersive media works – restoring /simulating each key project for viewing, alongside supporting historical documents. By linking each reconstructed experience with archival resources and historical metadata, we are creating an interactive platform that fosters access to information from formerly disconnected archival materials. This approach not only preserves XR history but also connects users with broader archival resources, promoting a holistic view of immersive media and the institutions that support it.

Our collaborations with the Computer History Museum and USC’s HMH Moving Image Archive underscore our commitment to building an interconnected archival framework. For example, the HMH Archive holds Morton Heilig’s original 3D films, widely considered the first VR films. At the time we contacted them, these films were not restored or publicly viewable. Our team worked with HMH to restore these films, which we then integrated into our VR experience of Heilig’s work. For the first time in decades, users were able to watch the 3D films as viewers did in the 1960s. Another example is our collaboration with the Computer History Museum (CHM), who gave us access to Ivan Sutherland’s materials, including both his original hardware and their complete oral history interviews. From access to their physical artifacts, we made 3D scans of Sutherland’s original head-mounted display and recreated Sutherland’s early demos from CHM’s materials, creating a virtual digital twin that can be viewed by anyone in the world.  
Again, wherever these seminal resources are currently housed, we work directly with that archive or museum to bridge their curation and restoration work to ours. We then prominently credit, feature, and link back to each of these institutions so that the public can access their work in turn. This approach models collaborative and multi-institutional resource-sharing approaches to archiving complex media. \cite{soccini2021museum} A core purpose of the Immersive Archive is to make XR’s early history accessible through immersive experiences, providing users with interactive educational resources. the Immersive Archive is a potential model within the archive network, connecting XR heritage with a global archival effort to ensure accuracy, integrity, and discoverability across digital media archives. By aligning with these goals, our proposal highlights the Immersive Archive as a shared contribution to the new media art archive community.

\section{Conclusion}
The Immersive Archive’s work with the Sensorama and Sutherland's Head-Mounted Display exemplifies the importance of preserving pioneering XR artifacts to understand the foundations of immersive media. Our approach to developing these simulations aims to offer an ever-evolving framework for future archival initiatives to preserve the rapidly changing field of XR. By recreating these experiences for contemporary VR headsets, we make them accessible to a new generation, underscoring the value of immersive history and its ongoing impact on society, technology, and media culture.

\section*{Acknowledgment}
The development team for this project consists of faculty, staff, and students from the School of Cinematic Arts (SCA) along with other schools on campus. The team is working in collaboration with SCA's HMH Foundation Moving Image Archive and The Computer History Museum. 

\bibliographystyle{IEEEtran}
\bibliography{references}

\begin{thebibliography}{10}
\providecommand{\url}[1]{#1}
\csname url@samestyle\endcsname
\providecommand{\newblock}{\relax}
\providecommand{\bibinfo}[2]{#2}
\providecommand{\BIBentrySTDinterwordspacing}{\spaceskip=0pt\relax}
\providecommand{\BIBentryALTinterwordstretchfactor}{4}
\providecommand{\BIBentryALTinterwordspacing}{\spaceskip=\fontdimen2\font plus
\BIBentryALTinterwordstretchfactor\fontdimen3\font minus \fontdimen4\font\relax}
\providecommand{\BIBforeignlanguage}[2]{{%
\expandafter\ifx\csname l@#1\endcsname\relax
\typeout{** WARNING: IEEEtran.bst: No hyphenation pattern has been}%
\typeout{** loaded for the language `#1'. Using the pattern for}%
\typeout{** the default language instead.}%
\else
\language=\csname l@#1\endcsname
\fi
#2}}
\providecommand{\BIBdecl}{\relax}
\BIBdecl

\bibitem{rheingold1991virtual}
H.~Rheingold, \emph{Virtual Reality: The Revolutionary Technology of Computer-Generated Artificial Worlds--And How It Promises to Transform Society}.\hskip 1em plus 0.5em minus 0.4em\relax New York: Simon \& Schuster, 1991.

\bibitem{messeri2024unreal}
\BIBentryALTinterwordspacing
L.~Messeri, \emph{In the Land of the Unreal: Virtual and Other Realities in Los Angeles}.\hskip 1em plus 0.5em minus 0.4em\relax Durham, NC: Duke University Press, 2024. [Online]. Available: \url{https://www.dukeupress.edu/in-the-land-of-the-unreal}
\BIBentrySTDinterwordspacing

\bibitem{soccini2017kusama}
A.~M. Soccini, ``Virtual environments before pixels: Yayoi kusama's impact on virtual reality,'' in \emph{Proceedings of the Conference Electronic Visualization and the Arts 2017}.\hskip 1em plus 0.5em minus 0.4em\relax BCS Learning \& Development, July 2017.

\bibitem{heilig_idea_books_2024}
{Morton Heilig Collection}, ``{Idea Books},'' Los Angeles, CA, USA.

\bibitem{10.1162/pres.1992.1.3.279}
\BIBentryALTinterwordspacing
M.~L. Heilig, ``El cine del futuro: The cinema of the future,'' \emph{Presence: Teleoperators and Virtual Environments}, vol.~1, no.~3, pp. 279--294, 08 1992. [Online]. Available: \url{https://doi.org/10.1162/pres.1992.1.3.279}
\BIBentrySTDinterwordspacing

\bibitem{VirtualRealityBeforeItHadThatName}
\BIBentryALTinterwordspacing
C.~H. Museum, ``Virtual reality before it had that name,'' 2019. [Online]. Available: \url{https://www.youtube.com/watch?v=Y2AIDHjylMI&ab_channel=ComputerHistoryMuseum}
\BIBentrySTDinterwordspacing

\bibitem{sutherlandhmd}
\BIBentryALTinterwordspacing
I.~E. Sutherland, ``A head-mounted three dimensional display,'' in \emph{Proceedings of the December 9-11, 1968, Fall Joint Computer Conference, Part I}, ser. AFIPS '68 (Fall, part I).\hskip 1em plus 0.5em minus 0.4em\relax New York, NY, USA: Association for Computing Machinery, 1968, p. 757–764. [Online]. Available: \url{https://doi.org/10.1145/1476589.1476686}
\BIBentrySTDinterwordspacing

\bibitem{ComputerHistoryMuseum}
{Computer History Museum}, ``Computer history museum,'' \url{https://www.computerhistory.org}.

\bibitem{chun2008enduring}
\BIBentryALTinterwordspacing
W.~H.~K. Chun, ``The enduring ephemeral, or the future is a memory,'' \emph{Critical Inquiry}, vol.~35, no.~1, pp. 148--171, 2008. [Online]. Available: \url{https://www.jstor.org/stable/10.1086/595632}
\BIBentrySTDinterwordspacing

\bibitem{soccini2021museum}
A.~M. Soccini and A.~M. Marras, ``Towards a standard approach for the design of a both physical and virtual museum,'' in \emph{2021 IEEE International Conference on Artificial Intelligence and Virtual Reality (AIVR)}, 2021, pp. 106--108.

\end{thebibliography}

\end{document}